\documentclass[%
 preprint, 
 amsmath,amssymb,
 aps, physrev,
]{revtex4-2}

\usepackage{graphicx}
\usepackage{dcolumn}
\usepackage{bm}
\usepackage[colorlinks=true, linkcolor=blue, citecolor=blue, urlcolor=blue]{hyperref}

\usepackage{url}

\DeclareUrlCommand\arxiv{%
  \urlstyle{rm}%
}

\begin{document}


\title{Searching for Axion-like particle Dark Matter with Time-domain Polarization:
Constraints from a protoplanetary disk}

\author{Kanako Narita}
\affiliation{Department of Astronomy, Graduate School of Science,
The University of Tokyo, 7-3-1 Hongo, Bunkyo-ku, Tokyo 113-0033, Japan}
\affiliation{National Astronomical Observatory of Japan, 2-21-1 Osawa, Mitaka, Tokyo 181-8588, Japan}

\author{Tomohiro Fujita}
\affiliation{Department of Physics, Ochanomizu University,
2-1-1 Otsuka, Bunkyo-ku, Tokyo 112-8610, Japan}
\affiliation{Kavli Institute for the Physics and Mathematics of the Universe (WPI), The University of
Tokyo Institutes for Advanced Study, The University of Tokyo, Chiba 277-8583, Japan}

\author{Ryo Tazaki}
\affiliation{Department of Earth Science and Astronomy, Graduate School of Arts and Sciences, The University of Tokyo, 3-8-1 Komaba, Meguro-ku, Tokyo 153-8902, Japan}

\author{Bunyo Hatsukade}
\affiliation{National Astronomical Observatory of Japan, 2-21-1 Osawa, Mitaka, Tokyo 181-8588, Japan}
\affiliation{Graduate University for Advanced Studies, SOKENDAI, Osawa, Mitaka, Tokyo 181-8588, Japan}
\affiliation{Department of Astronomy, Graduate School of Science, The University of Tokyo, 7-3-1 Hongo, Bunkyo-ku, Tokyo 133-0033, Japan}

\date{\today}

\begin{abstract}
Axion-like particles (ALPs) can induce a rotation of polarization plane effect that rotates the polarization angle of light, offering a probe of ultralight dark matter.
We analyze archival near-infrared polarimetric data of the protoplanetary disk (PPD) around HD~163296. Building on previous snapshot and multi-epoch studies of protoplanetary-disk polarization, we perform a dedicated time-series analysis of archival near-infrared polarimetric data of HD 163296. 
The resulting six-epoch time series is consistent with a constant polarization angle within the measurement uncertainties, while being sensitive to timescales of $\sim$170–400 days. The typical polarization angle uncertainties are $1.6^\circ$–$6.4^\circ$, and are partly driven by multiple scattering in the optically thick disk, which broadens the intrinsic polarization angle distribution and introduces additional dispersion in the representative polarization angle.
Based on these data, we derive the first upper limits on the ALP–photon coupling from PPD polarization variability, $g_{a\gamma} \lesssim 7.5 \times 10^{-12}\,(m_a / 10^{-22}\,\mathrm{eV})\,\mathrm{GeV}^{-1}$. 
Finally, we show that, if future observations can achieve a total polarization-angle uncertainty of $\sigma_\chi\sim0.1^\circ$, including systematic effects, time-domain polarimetry of protoplanetary disks could rival or surpass existing astrophysical constraints on the ALP--photon coupling in the relevant mass range.
\end{abstract}

\maketitle


\section{\label{sec:intro}Introduction}

The identification of the nature of dark matter has remained a key open problem in both physics and astronomy for decades. While the cold dark matter (CDM) paradigm successfully reproduces the large-scale structure of the Universe, it faces several tensions with observations on galactic scales, such as the core--cusp problem \citep{2015PNAS..11212249W}. As a potential resolution to these discrepancies, dark matter with an extremely small mass ($m \sim 10^{-22}~\mathrm{eV}$) has attracted considerable attention~\cite{2000PhRvL..85.1158H,2017PhRvD..95d3541H}. At such ultralight masses, quantum mechanical effects become significant even on astrophysical scales, leading to the suppression of structure formation on small scales.

Among the many dark matter candidates, axion-like particles (ALPs) are a particularly promising option to account for these discrepancies \cite{1983PhLB..120..127P,1983PhLB..120..133A,1983PhLB..120..137D}. 
Among the many dark matter candidates, axion-like particles (ALPs) are a particularly promising option. 
ALPs are pseudoscalar fields that can couple to photons through the interaction $g_{a\gamma}\phi F_{\mu\nu}\tilde{F}^{\mu\nu}/4$.
The original axion was introduced as a pseudo-Nambu--Goldstone boson associated with the Peccei--Quinn mechanism to solve the strong CP problem in quantum chromodynamics (QCD)~\cite{1977PhRvL..38.1440P,1977PhRvD..16.1791P,1978PhRvL..40..223W,1978PhRvL..40..279W}, and ALPs are also predicted by string theories~\cite{1984PhRvD..30..272W,2010PhRvD..81l3530A,2012JHEP...10..146C}. Unlike the QCD axion, the mass and coupling constants of ALPs are not necessarily related, allowing for a wide range of phenomenological possibilities. In particular, ALPs with extremely small masses ($m \sim 10^{-22}~\mathrm{eV}$) can behave as ultralight bosonic fields and constitute so-called fuzzy dark matter, which suppresses small-scale structure formation through quantum pressure effects~\cite{2000PhRvL..85.1158H,2014NatPh..10..496S}. This scenario not only provides a viable dark matter candidate, but also offers a natural resolution to the small-scale problems inherent in the standard CDM framework. Here, we assume that dark matter, including ALPs, is of the Axion Dark Matter (ADM) type.
It should be noted that this mass range is already subject to important cosmological and astrophysical constraints. Because ultralight axions suppress the growth of small-scale structure through quantum-pressure effects, observations of the Lyman-$\alpha$ forest \cite{2021PhRvL.126g1302R}, dwarf galaxies \cite{2017PhRvD..95d3541H, 2019PhRvL.123e1103M}, pulsar timing arrays \cite{2023PhRvL.131q1001S}, and black-hole superradiance \cite{2011PhRvD..83d4026A} place constraints on the allowed axion mass and its contribution to the total dark-matter density. These probes primarily test the gravitational effects of ultralight dark matter. In contrast, the present work probes the axion--photon coupling through polarization-angle oscillations. Our approach is therefore complementary to existing cosmological and astrophysical searches, providing an independent test of the interaction between ALPs and photons.

ADM has been searched for across a broad mass range using a variety of optical and astrophysical approaches, yet there remain mass ranges where current methods have limited sensitivity. Two main observational strategies have been employed to search for ADM. The first is the conversion between axions and photons in the presence of magnetic fields, which has been exploited in laboratory experiments such as axion helioscopes~\cite{2017NatPh..13..584C,2014JInst...9.5002A} and light-shining-through-a-wall setups~\cite{2010PhLB..689..149E}. Astrophysical observations, including studies of SN1987A~\cite{2015JCAP...02..006P} and distant quasars~\cite{2012JCAP...07..041P}, have also been used to search for signatures of this conversion. To date, no definitive signal has been detected (though see Refs.~\cite{2014JCAP...09..026A,2013PhRvD..87c5027M,2017PhRvD..96e1701K} for tentative hints). Because the conversion probability depends on the coupling constant $g_{a\gamma}$ and on the strength and morphology of intervening magnetic fields, these astrophysical constraints typically set upper limits of $g_{a\gamma} \lesssim 10^{-11}~\mathrm{GeV}^{-1}$ for $m \lesssim 10^{-14}~\mathrm{eV}$, but remain subject to uncertainties in cosmic magnetic field properties.

The second approach is to search for the rotation of the polarization plane induced by an axion background~\cite{1990PhRvD..41.1231C,1992PhLB..289...67H}. 
If ADM exists, its coupling to photons can lead to a rotation of the polarization angle of light, depending on the variation of the ALP field between the emission and detection points (see Section~\ref{sec:adm} for details). In previous applications, this method has been applied to polarization maps of the cosmic microwave background (CMB), where the rotatio polarization plane induced by axion-like or axion dark matter is probed through map-based statistical analyses rather than time-domain measurements. In particular, the spatial power spectrum of the polarization rotation angle and its impact on the CMB polarization power spectra, including the depolarization (washout) of the primordial E modes, are evaluated and compared with observations to constrain the ADM–photon coupling and the particle mass \citep{2019PhRvD.100a5040F}. While the majority of previous rotation of polarization plane searches have been limited to static, map-based analyses, recent efforts have extended to the time domain \citep[e.g.,][]{2019JCAP...02..059I,2019PhRvD.100f3515C}. These studies target bright, polarized astrophysical sources such as Tau A and pulsars, aiming to detect periodic or otherwise temporally varying polarization rotations \citep{2024PhRvD.110f3013A}.
These studies target a variety of bright polarized astrophysical sources, including blazars \citep{2019JCAP...02..059I}, pulsars and fast radio bursts (FRBs) \citep{2019PhRvD.100f3515C}, and the Crab Nebula (Tau A) \citep{2024PhRvD.110f3013A}, aiming to detect periodic or otherwise temporally varying polarization rotations.

By monitoring these sources over extended periods, time-domain analyses have, in certain cases, exceeded the sensitivity of static approaches, with recent pulsar timing array (PTA) observations  providing the most stringent constraints to date on specific regions of the ADM parameter space \citep{2024arXiv241202229X}.

Protoplanetary disks (PPDs) are flattened, optically thick composed of gas and dust surrounding young stars, where planets are thought to form. The optical thickness arises primarily from dust absorption and scattering. In optical and near-infrared bands, the disk is visible in scattered stellar light by dust grains in the disk surface regions, which produces a well-understood linear polarization pattern at the source \citep[e.g.,][]{2023ASPC..534..605B}. This predictable geometry allows the observed polarization angle to serve as a sensitive probe of any rotation caused by the rotation polarization plane during photon propagation. PPDs offer a large ensemble of spatially resolved, polarized sources observed over multiple epochs, providing numerous measurements of polarization angles. This makes them ideal targets for probing potential time-dependent rotation polarization plane effects predicted by ADM--photon interactions, while also enhancing the statistical sensitivity of such searches~\citep{2020PhLB..80335288C}. Using this method, a study analyzing snapshot polarization images of PPDs has placed a constraint on the coupling constant of $g_{a\gamma} < 5 \times 10^{-13} \left( \frac{m}{10^{-22}~\mathrm{eV}} \right) \ \mathrm{GeV}^{-1}$~\citep{2019PhRvL.122s1101F}. 
However, in reality protoplanetary disks often deviate from strict axisymmetry: planet formation, magnetic fields, spiral density waves, and gravitational instabilities can all introduce intrinsic departures from centrosymmetric polarization pattern \citep[e.g.,][]{2023ASPC..534..605B}. 
While such effects do not necessarily invalidate snapshot-based polarization analyses—particularly when deriving conservative constraints, they indicate that care must be taken when discussing the detection of temporal variations under an axisymmetric assumption \citep{2019PhRvL.122s1101F}.
More recently, Davydov and Libanov
\cite{2024PhRvD.110j3022D}
used multi-epoch ALMA polarization observations of the
protoplanetary disk HL Tau and obtained stronger constraints
on the ALP--photon coupling. Their analysis was based on
polarization-angle differences between two observing epochs.


In this work, we adopt a complementary time-series approach based on near-infrared scattered-light polarimetry of the protoplanetary disk HD~163296. By modeling the temporal evolution expected from ALP-induced polarization rotation and searching directly for sinusoidal modulation in the polarization-angle time series, our method preserves the connection between oscillation period and ALP mass while naturally incorporating the temporal sampling and cadence of the observations. Because the analysis relies on relative temporal variations within the same system rather than on the absolute polarization pattern, it remains robust even when the disk exhibits intrinsic asymmetries. This provides a complementary avenue for ALP searches that is less sensitive to the detailed astrophysical morphology of individual PPDs.

The structure of this paper is as follows.
Section~\ref{sec:intro} introduces the scientific background and motivation, emphasizing the potential of PPDs for probing ultralight axion-like dark matter.
Section~\ref{sec:adm} summarizes the theoretical framework for ALP-induced polarization rotation.
Section~\ref{sec:methods} explains the methodology for estimating polarization angles and their uncertainties from the polarization maps, as well as for deriving upper limits on the coupling constant.
Section \ref{sec:results} presents the results of our analysis, including the constraints on the axion--photon coupling from the PPD around HD 163296 and forecasts of the observational requirements for future polarization-angle precision.
Section \ref{sec:discussion} discusses observational prospects, limitations of joint analyses, and potential systematic effects.
Finally, Section \ref{sec:conclusion} summarizes our findings and outlines future directions.

\section{Theoretical framework of ADM-induced polarization rotation}
\label{sec:adm}
In this section, we outline the theoretical framework of ADM-induced polarization rotation.
The rotation angle of the photon polarization plane is proportional to the difference 
in the ALP field values between the source at the emission time $(t_{\rm emit},\bm x_{\rm emit})$ and the observer at the observation time $(t_{\rm obs},\bm x_{\rm obs})$ \citep{1992PhLB..289...67H, 2019PhRvL.122s1101F}:
\begin{equation}
\chi(t_{\rm obs},x_{\rm obs},t_{\rm emit},x_{\rm emit}) = -\frac{g_{a\gamma}}{2} \left[ \phi(t_{\rm obs},\bm x_{\rm obs}) - \phi(t_{\rm emit},\bm x_{\rm emit}) \right],
\label{eq:theta_general}
\end{equation}
where $g_{a\gamma}$ is the ADM–photon coupling constant.
The local ADM field in our Milky Way Galaxy can be approximated by \cite{2023PhRvD.108i2010N}
\begin{equation}
\phi(t,\bm x) = \phi_0 \cos(m_a t + \delta(\bm x)),
\end{equation}
where $\phi_0$ is the averaged amplitude, $\delta(\bm x)$ is the phase depending on the position, and $m_a$ is the ADM mass.
Substituting this into Equation~\eqref{eq:theta_general} and rearranging, we obtain
\begin{equation}
\chi = g_{a\gamma} \,\phi_0\, \sin\Xi\,
\sin\!\left( m_a t_{\rm obs} + \Xi + \delta_{emit} \right),
\end{equation}
where $\Xi \equiv \frac{1}{2}\left[m_a (t_{\rm obs}-t_{\rm emit})+\delta_{\rm obs}-\delta_{\rm emit}\right]$ is constant because $t_{\rm obs}-t_{\rm emit} = \frac{d}{c}$, which corresponds to the fixed light-travel time from the source to the observer, with $d$ being the source–observer distance. In this form, the rotation angle is expressed as a time oscillation with a coefficient of $\sin\Xi$.
From the local dark matter density, $\rho_{\rm DM} \simeq 0.3~\mathrm{GeV\,cm^{-3}}$ \citep{2014JPhG...41f3101R},  
the amplitude can be estimated as $\phi_0^2 \simeq 2\rho_{\rm DM} / m_a^2$.  
For convenience, we define
\[
g_{12} \equiv \frac{g_{a\gamma}}{10^{-12}~\mathrm{GeV}^{-1}}, \quad
m_{22} \equiv \frac{m_a}{10^{-22}~\mathrm{eV}}.
\]
The rotation angle can then be written as
\begin{equation}
\chi(t,T) \approx 2\times 10^{-2}\,
\sin\Xi\,\sin(m_a t_{\rm obs} + \Xi + \delta_{\rm emit})\
g_{12}\,m_{22}^{-1}.
\label{eq:theta_amp}
\end{equation}

In practical observations, 
$\Xi$ is dominated by the term 
$m_a (t_{\rm obs} - t_{\rm emit})$, 
which becomes much greater than unity for the ALP masses considered here.
Because $\Xi \gg 1$, the phase is effectively randomized, 
and $\sin\Xi$ can be approximated as a random variable 
with mean zero and rms $1/\sqrt{2}$.
Applying this to the amplitude yields
\begin{equation}
\chi(t_{\rm obs},T) \simeq 1.4\times 10^{-2}\,
\sin(m_a t_{\rm obs} + \mathrm{const.})\,
g_{12}\,m_{22}^{-1},
\label{eq:theta_final}
\end{equation}
indicating that the polarization angle exhibits a sinusoidal time variation with angular frequency $\omega = m_a$,  
and that the effective amplitude is proportional to $g_{a\gamma}/m_a$.
Equation~\eqref{eq:theta_final} assumes that the local ALP field remains coherent over the duration of the observations. For virialized Galactic dark matter with a characteristic velocity dispersion $v\sim10^{-3}c$, the coherence time is approximately $t_{\rm coh}\sim(m_a v^2)^{-1}$, which is generally much longer than typical observational timescales. The adiabatic approximation is also well satisfied because the ALP oscillation frequency is much smaller than the photon frequency. Therefore, the ALP field can be treated as a coherent sinusoidal background over the duration of the observations.

\begin{figure}[t]
\centering
\includegraphics[width=0.8\textwidth]{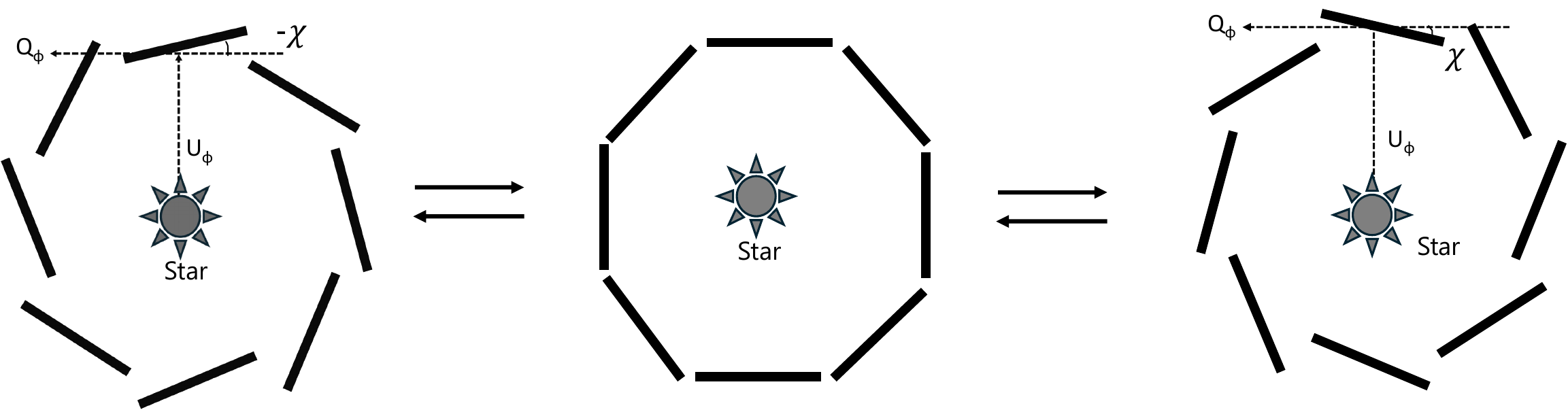}
\caption{
Schematic illustration of the polarization geometry in a protoplanetary disk (PPD) and the expected effect of ALP-induced rotation of polarization plane.
The central panel shows the intrinsic azimuthal polarization pattern produced by scattered light on the disk surface.
Polarization is analyzed using local Stokes parameters $Q_\phi$ and $U_\phi$ defined in the azimuthal frame; the left and right panels illustrate the corresponding polarization orientations.
In the presence of ALP induced rotation of polarization plane, the polarization plane is rotated by an angle $\chi$, which oscillates in time, producing a time-varying polarization angle as described by Equation~\ref{eq:theta_final}.
}
\label{fig:schematic_ppd}
\end{figure}

In previous work based on single-epoch (snapshot) polarization images of PPDs~\cite{2019PhRvL.122s1101F},
the time-dependent factor $\sin(m_a t + \mathrm{const})$ could not be observationally
resolved, because snapshot images do not provide temporal sampling of the modulation.
As a result, the amplitude could only be estimated by adopting its rms expectation
value of $1/\sqrt{2}$. In contrast, the multi-epoch observations used in this study
allow us to sample the temporal evolution predicted by ~\eqref{eq:theta_final} and to search directly
for the sinusoidal modulation without relying on this rms-based approximation.
This represents a key methodological improvement over previous snapshot-based analyses.
In the context of PPDs, the intrinsic polarization geometry is well understood:
scattered light from submicron-sized dust grains on the disk surface produces polarization vectors 
that are nearly azimuthal, i.e., perpendicular to the radial direction from the central star 
\citep[e.g.,][]{2011ApJ...729L..17H}. 
This provides a well-defined reference orientation for the polarization angle at each epoch.
Thus, since axion-induced polarization rotation arises as a propagation effect, it is expected to act uniformly across the disk rather than producing spatially varying deviations from the azimuthal pattern. In this work, we thus focus on the disk-integrated polarization angle by computing an inverse-variance weighted mean angle for each epoch, and search for temporal variations in this spatially averaged quantity, as would be induced by an axion-like field according to Equation~\eqref{eq:theta_final}.
Figure~\ref{fig:schematic_ppd} illustrates this concept, where the scattered light from the disk surface 
produces azimuthal polarization vectors, and a small additional rotation $\chi$ represents the 
possible rotation of pplarization plane signal to be searched for.

\section{Methods}
\label{sec:methods}
This section details our approach for estimating polarization angles and their measurement uncertainties, followed by the framework used to compute upper limits on temporal polarization rotation in the presence of axion-like dark matter.
\subsection{Estimation of polarization angles and uncertainties}
We first derive the azimuthal Stokes parameters $Q_\phi$ and $U_\phi$ by rotating
the linear Stokes parameters $Q$ and $U$ into a polar coordinate frame centered
on the star \citep{2019ApJ...872..122M, 2020A&A...633A..63D}:
\begin{align}
Q_\phi &= -Q \cos(2\phi) - U \sin(2\phi), \nonumber \\
U_\phi &= \phantom{-}Q \sin(2\phi) - U \cos(2\phi),
\label{eq:qphi_uphi}
\end{align}
where $\phi$ is the position angle of each pixel measured from north through east.
In this convention, $Q_\phi > 0$ corresponds to an azimuthal (tangential)
polarization pattern, while $U_\phi$ represents deviations from perfect
azimuthal symmetry.
We then construct polarized intensity maps as
\begin{equation}
P = \sqrt{Q_\phi^2 + U_\phi^2}.
\label{eq:polarized_intensity}
\end{equation}
The noise level $\sigma$ is estimated from emission-free regions of the
$U_\phi$ map. Pixels satisfying $R > 5\sigma$ are retained and grouped into
$3\times 3$ pixel blocks, each corresponding to one independent beam. A circular region of radius 22 pixels centered on the stellar position is additionally masked to remove the central area strongly affected by stellar emission.
At the distance of HD~163296, the 22- and 70-pixel cuts correspond to
projected radii of approximately 27 and 87 au, respectively, for the
VLT/SPHERE/IRDIS pixel scale of $12.25$ mas pixel$^{-1}$
\citep{2023A&A...680A.114R}.
In addition, to confine the analysis to the observed extent of the
scattered-light disk and suppress spurious high-significance features
near the edge of the field, all pixels at projected radii exceeding
$r_{\rm out}=70$ pixels from the stellar position are excluded.
The adopted value of $r_{\rm out}$ encompasses the full bright
scattering ring in all epochs while avoiding isolated background
fluctuations. We have verified that moderate variations of
$r_{\rm out}$ within a reasonable range do not qualitatively
affect the derived upper limits.

For each block $b = 1,\dots,N_{\rm block}$, corresponding to an individual $3\times3$ pixel block, let its valid pixels be indexed by $i = 1,\dots,N_{{\rm pix},b}$.
Here, the index $b$ labels individual $3\times3$ blocks, and $N_{{\rm pix},b}$ denotes the number of finite pixels within each block, which may vary due to masking and data selection.
The block–averaged Stokes parameters are
\begin{equation}
\bar{Q}_{\phi,b}
=
\frac{1}{N_{{\rm pix},b}}
\sum_{i=1}^{N_{{\rm pix},b}} Q_{\phi,i},
\qquad
\bar{U}_{\phi,b}
=
\frac{1}{N_{{\rm pix},b}}
\sum_{i=1}^{N_{{\rm pix},b}} U_{\phi,i}.
\label{eq:def_avg_QU}
\end{equation}

Assuming a uniform rms noise $(\sigma_Q, \sigma_U)$ across the image, the
uncertainties of the block-averaged Stokes parameters follow from standard
variance propagation:
\begin{equation}
\sigma_{\bar{Q},b}^{2}
=
\frac{\sigma_Q^{2}}{N_{{\rm pix},b}},
\qquad
\sigma_{\bar{U},b}^{2}
=
\frac{\sigma_U^{2}}{N_{{\rm pix},b}}.
\label{eq:sigma_avg_QU_definition}
\end{equation}

From these block-averaged values, the polarization angle was computed as

\begin{equation}
\chi_b
=
\frac{1}{2}
\arctan\!\left(
\frac{\bar{U}_{\phi,b}}{\bar{Q}_{\phi,b}}
\right).
\label{eq:chi_b_definition}
\end{equation}

which follows from the standard definition of the polarization angle.

The uncertainty of $\chi$ was obtained via linear error propagation.
For a function
\begin{equation}
\chi_b = f(\bar{Q}_{\phi,b}, \bar{U}_{\phi,b}),
\label{eq:chi_b_definition2}
\end{equation}

the variance is given by
\begin{equation}
\sigma_{\chi,b}^2
=
\left( \frac{\partial f}{\partial \bar{Q}_{\phi,b}} \right)^2
\sigma_{\bar{Q},b}^{\,2}
+
\left( \frac{\partial f}{\partial \bar{U}_{\phi,b}} \right)^2
\sigma_{\bar{U},b}^{\,2},
\label{eq:error_propagation_general}
\end{equation}

where we have assumed that the uncertainties in $\bar{Q}_\phi$ and $\bar{U}_\phi$ are uncorrelated.

The required partial derivatives are
\begin{equation}
\frac{\partial \chi_b}{\partial \bar{Q}_{\phi,b}}
= -\frac{1}{2}\,
\frac{\bar{U}_{\phi,b}}{\bar{Q}_{\phi,b}^{\,2} + \bar{U}_{\phi,b}^{\,2}},
\qquad
\frac{\partial \chi_b}{\partial \bar{U}_{\phi,b}}
= \frac{1}{2}\,
\frac{\bar{Q}_{\phi,b}}{\bar{Q}_{\phi,b}^{\,2} + \bar{U}_{\phi,b}^{\,2}},
\label{eq:partials_chi}
\end{equation}

which, inserted into Equation~\eqref{eq:error_propagation_general}, give
\begin{align}
\sigma_{\chi,b}^2
&=
\left[
    -\frac{1}{2}
    \frac{\bar{U}_{\phi,b}}{\bar{Q}_{\phi,b}^{\,2} + \bar{U}_{\phi,b}^{\,2}}
\right]^2
\sigma_{\bar{Q},b}^{\,2}
+
\left[
    \frac{1}{2}
    \frac{\bar{Q}_{\phi,b}}{\bar{Q}_{\phi,b}^{\,2} + \bar{U}_{\phi,b}^{\,2}}
\right]^2
\sigma_{\bar{U},b}^{\,2}
\label{eq:sigma_chi_expand}
\\[4pt]
&=
\frac{1}{4}
\frac{
    \bar{Q}_{\phi,b}^{\,2}\,\sigma_{\bar{U},b}^{\,2}
    +
    \bar{U}_{\phi,b}^{\,2}\,\sigma_{\bar{Q},b}^{\,2}
}{
    \left(\bar{Q}_{\phi,b}^{\,2} + \bar{U}_{\phi,b}^{\,2}\right)^2
}.
\label{eq:sigma_chi_expand2}
\end{align}
Taking the square root, the uncertainty of the polarization angle becomes
\begin{equation}
\sigma_{\chi,b} =
\frac{1}{2}
\frac{
\sqrt{
    \bar{Q}_{\phi,b}^{\,2}\,\sigma_{\bar{U},b}^{\,2}
    +
    \bar{U}_{\phi,b}^{\,2}\,\sigma_{\bar{Q},b}^{\,2}
}}
{\bar{Q}_{\phi,b}^{\,2} + \bar{U}_{\phi,b}^{\,2}}.
\label{eq:evpa_block_sigma_chi}
\end{equation}

This block-averaging procedure effectively reduces pixel–pixel correlations by
treating each $3\times 3$ pixels region as one independent beam, thereby providing a
more stable estimate of the per–epoch polarization angle and its statistical
uncertainty.

After computing the polarization angles and their uncertainties for all
independent blocks, we derived the disk–integrated polarization angle for
each epoch. In studies of interstellar polarization, the measurement
uncertainties are often relatively homogeneous across sight lines, and
therefore polarization angles are frequently averaged without applying
explicit statistical weighting \citep[e.g.,][]{2020ApJ...899...28D}.
However, Appendix~C of \citet{2020ApJ...899...28D} demonstrates that
axial (180$^\circ$-periodic) angles can be combined within the framework of
circular statistics by assigning each doubled-angle vector a length
proportional to a chosen weight. In our application, where the per-block
uncertainties vary substantially across the disk, such a weighted formulation
is essential. We therefore adopt this generalized approach by letting each
block contribute a doubled-angle vector whose length is given by the
inverse-variance weight $w_b = 1/\sigma_{\chi,b}^{2}$. This provides a
statistically consistent extension of standard circular averaging and leads
naturally to the weighted axial mean described below.
Because the polarization angle is an axial (180$^\circ$-periodic) quantity,
the averaging must be performed in the doubled-angle space ($2\theta$)
\citep{MardiaJupp2000}. The weighted mean angle is therefore
\begin{equation}
\bar{\chi}
=
\frac{1}{2}\,
\arctan2\!\left(
    \sum_b w_b \sin 2\chi_b,\,
    \sum_b w_b \cos 2\chi_b
\right),
\label{eq:weighted_mean_evpa}
\end{equation}
where $w_b = 1/\sigma_{\chi,b}^2$ are inverse-variance weights.

Here, $\arctan2(y,x)$ denotes the quadrant-preserving inverse tangent,
defined by
\begin{equation}
\arctan2(y,x) =
\begin{cases}
\arctan\!\left(\dfrac{y}{x}\right), & x > 0, \\[8pt]
\arctan\!\left(\dfrac{y}{x}\right) + \pi, & x < 0,\; y \ge 0, \\[8pt]
\arctan\!\left(\dfrac{y}{x}\right) - \pi, & x < 0,\; y < 0, \\[8pt]
+\dfrac{\pi}{2}, & x = 0,\; y > 0, \\[8pt]
-\dfrac{\pi}{2}, & x = 0,\; y < 0, \\[8pt]
\text{undefined}, & x = 0,\; y = 0.
\end{cases}
\label{eq:atan2_def}
\end{equation}
Unlike the conventional $\arctan(y/x)$, which is restricted to the
range $(-\pi/2,\, \pi/2)$ and therefore loses quadrant information,
$\arctan2(y,x)$ returns an angle in the full range $(-\pi, \pi]$,
ensuring that the correct angular branch is used when averaging axial
(180$^\circ$-periodic) quantities such as the polarization angle.

The standard deviation of the resulting distribution of representative
polarization angles is adopted as conservative estimate of the
uncertainty associated with the representative angle.
In practice, it is computed as \citep{MardiaJupp2000}
\begin{equation}
\sigma_{\rm axial}
=
\frac{1}{2}\,
\sqrt{-2\ln R},
\qquad
R = \frac{\sqrt{\left(\sum_b w_b \cos 2\chi_b\right)^2
+
\left(\sum_b w_b \sin 2\chi_b\right)^2}}{\sum_b w_b}.
\label{eq:sigma_axial}
\end{equation}
Here, $\sigma_{\rm axial}$ quantifies the scatter of the polarization
angles among the spatial blocks within a given epoch. It therefore represents
the intrinsic dispersion of the measured angles across the disk rather than
the formal standard error of the mean polarization angle. The latter would be
smaller by approximately a factor of $\sqrt{N_{\rm eff}}$, where
$N_{\rm eff}$ is the number of statistically independent blocks. Throughout
this work, we adopt $\sigma_{\rm axial}$ as a conservative uncertainty estimate
because it naturally incorporates both spatial variations in the polarization
pattern and measurement uncertainties across the disk.
This procedure correctly accounts for both the angular periodicity of the
polarization angle and the spatially varying uncertainties across the disk,
providing a statistically robust estimate of the epoch-averaged polarization
angle and its corresponding uncertainty.


\subsection{Estimation of upper limits}

To constrain possible polarization-angle modulations induced by ALPs, 
we quantify the maximum sinusoidal amplitude consistent with the data.  
Because the number of epochs is limited in the dataset used in this study, a full periodic search (e.g., least-squares spectral analysis) 
is not statistically meaningful.  
Instead, we derive frequentist upper limits based on Monte Carlo (MC) simulations that 
faithfully reproduce the actual sampling cadence and measurement uncertainties.

For each target, we first construct the polarization angle time series $\{\chi_i\}$ and corresponding 
$1\sigma$ uncertainties $\{\sigma_i\}$ , where the index $i$ labels individual observing epochs.
Under the null hypothesis of no intrinsic polarization angle variation, we generate 
$N_{\rm MC}=10^5$ mock time series $\{\chi_i^{(k)}\}$ by drawing Gaussian noise realizations 
with standard deviations $\sigma_i$ at the observed epochs.  
Each realization is fitted with a sinusoidal model,
\begin{equation}
  \chi(t) \;=\; A\,\sin\!\left(2\pi f t + \phi\right) + \chi_0,
  \label{eq:sin_model}
\end{equation}
where $A$ is the modulation amplitude, $f$ is the trial frequency corresponding to an 
ALP mass $m_a = 2\pi f$, $\phi$ is the phase, and $\chi_0$ is the mean polarization angle.  
We minimize the $\chi^2$ statistic over $(A,\,\phi,\,\chi_0)$ for each realization to obtain 
the best-fit amplitude $A^{(k)}$.

The resulting distribution $\{A^{(k)}\}$ represents the noise-only expectation for each frequency.  
From this distribution, we define the 95\% confidence-level (CL) upper limit $A_{95}(f)$ by
\begin{equation}
  \mathbb{P}\bigl(A^{(k)} < A_{95}(f)\bigr) = 0.95 .
  \label{eq:A95_def}
\end{equation}
Applying the same fitting procedure to the observed data shows that all trial frequencies 
are consistent with the null hypothesis within the 95\% CL.

To express the theoretical amplitude predicted by Equation~\ref{eq:theta_final} in degrees,
we first note that Equation \ref{eq:theta_final} gives the axion-induced polarization rotation
with an amplitude in radians,
\begin{equation}
A_{\rm theory}({\rm rad})
= 1.4\times10^{-2}\, g_{12}\, m_{22}^{-1}.
\label{eq:atheory_rad}
\end{equation}
Using the standard conversion $1~{\rm rad} = 180/\pi~{\rm deg}$,
the corresponding amplitude in degrees becomes
\begin{equation}
A_{\rm theory}({\rm deg})
= 0.80^\circ\, g_{12}\, m_{22}^{-1}.
\label{eq:atheory_deg}
\end{equation}
Requiring that the observed 95\% upper limit $A_{95}$ satisfies
$A_{95} \ge A_{\rm theory}$, the axion--photon coupling is constrained as

\begin{equation}
  g_{12}
  \;<\;
  1.25\,
  \left(\frac{A_{95}}{1^{\circ}}\right)\,
  m_{22},
  \qquad\text{(for $\rho_{\rm DM}=0.3~\mathrm{GeV\,cm^{-3}}$)}.
  \label{eq:gagg_limit_rho03}
\end{equation}

\section{Data and Results}
\label{sec:results}
\subsection{Observational Data}
The target of this study, the protoplanetary disk around HD~163296 (R.A. = 17:56:21.29, decl. = -21:57:21.87), is one of the sources included in larger SPHERE/IRDIS survey of 29 protoplanetary disks obtained in the Ks band using the star-hopping mode \cite{2023A&A...680A.114R}.  
The observations were carried out under six ESO programs
(0103.C-0470, 105.209E, 105.20HV, 106.21HJ, 108.22EE; PI: M.~Benisty,
and 105.20JB; PI: M.~Keppler).  
These data were taken in DPI mode with pupil tracking, allowing simultaneous
total-intensity and polarized-light imaging.  
In the star-hopping strategy, each disk-hosting target is paired with a
well-matched reference star to obtain quasi-simultaneous wavefront calibration,
thereby improving speckle subtraction and the detection of circumstellar
structures \citep{2021A&A...648A..26W}.
Among the survey targets, a protoplanetary disk around HD~163296 offers the richest time-domain sampling, providing six epochs suitable for our polarization angle variability analysis.  The six SPHERE/IRDIS epochs used in this study were obtained on  
2021 April 6, 2021 June 3, 2021 September 9,  
2021 September 26, 2022 June 11, and 2022 July 7.
A comprehensive description of the survey design, observing setup, and data
reduction procedures is presented in \cite{2023A&A...680A.114R}. 
The stellar properties of HD~163296 are taken from \cite{2023A&A...680A.114R} and summarized in Table~\ref{tab:stellar_properties}.

\begin{table}
\caption{Stellar properties and observing log of HD~163296 adopted in this work \cite{2023A&A...680A.114R}.}
\label{tab:stellar_properties}
\centering
\begin{tabular}{lc}
\hline
Property & Value \\
\hline
Associated molecular cloud & None \\
Distance & $101.0^{+0.4}_{-0.4}$ pc \\
Stellar luminosity & $15.9^{+0.3}_{-0.3}\,L_\odot$ \\
Stellar mass & $2.0\,M_\odot$ \\
Age & 10--12 Myr \\
\hline
Date (UTC) & $t_{\rm exp}$ (s) \\
\hline
2021-04-06 & 384 \\
2021-06-03 & 2048 \\
2021-09-09 & 1536 \\
2021-09-26 & 512 \\
2022-06-11 & 3072 \\
2022-07-07 & 3072 \\
\hline
\end{tabular}
\end{table}

\subsection{The polarization angle measurements and constraints on coupling constant from the protoplanetary disk around HD 163296} 
\begin{figure}[t]
\centering
\includegraphics[width=0.65\textwidth]{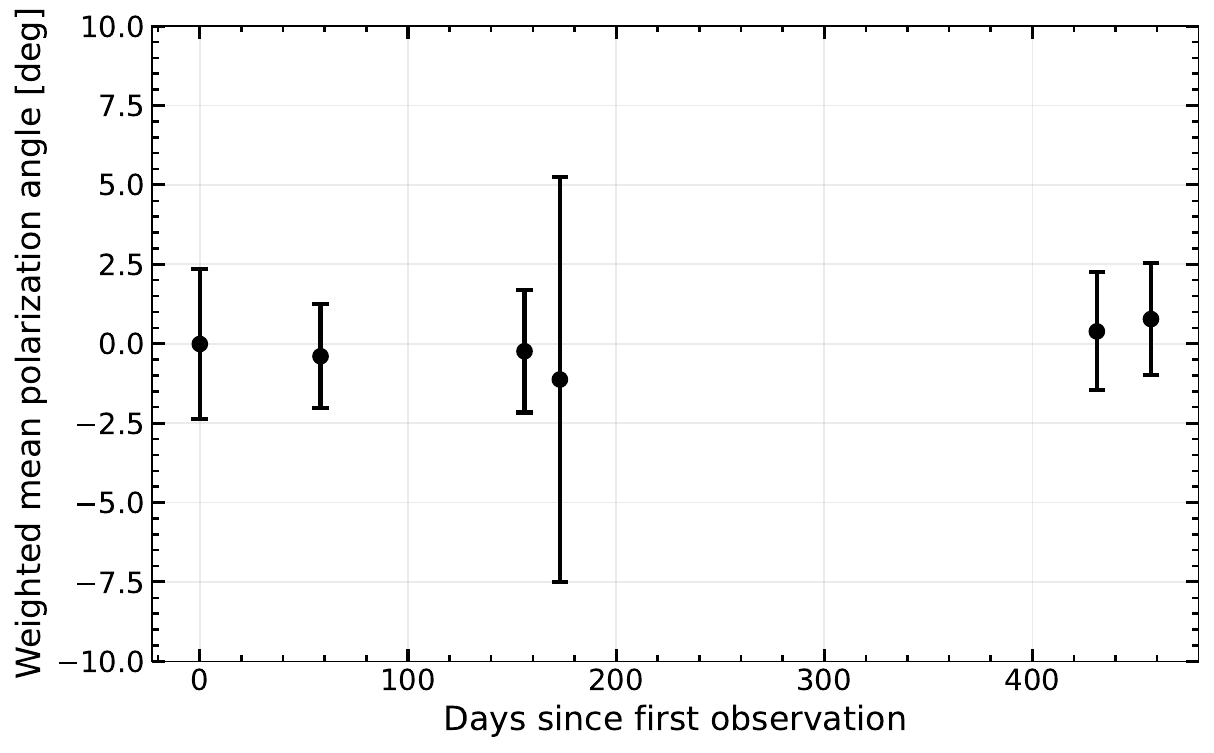}
\caption{The mean value of the polarization angle of HD~163296 as a function of days since 2021 April 6. 
Error bars represent the $1\sigma$ statistical uncertainties derived from the Stokes $Q_\phi$ and $U_\phi$ maps.}
\label{fig:evpa_HD163296}
\end{figure}

\begin{figure}[t]
\centering
\includegraphics[width=0.8\textwidth]{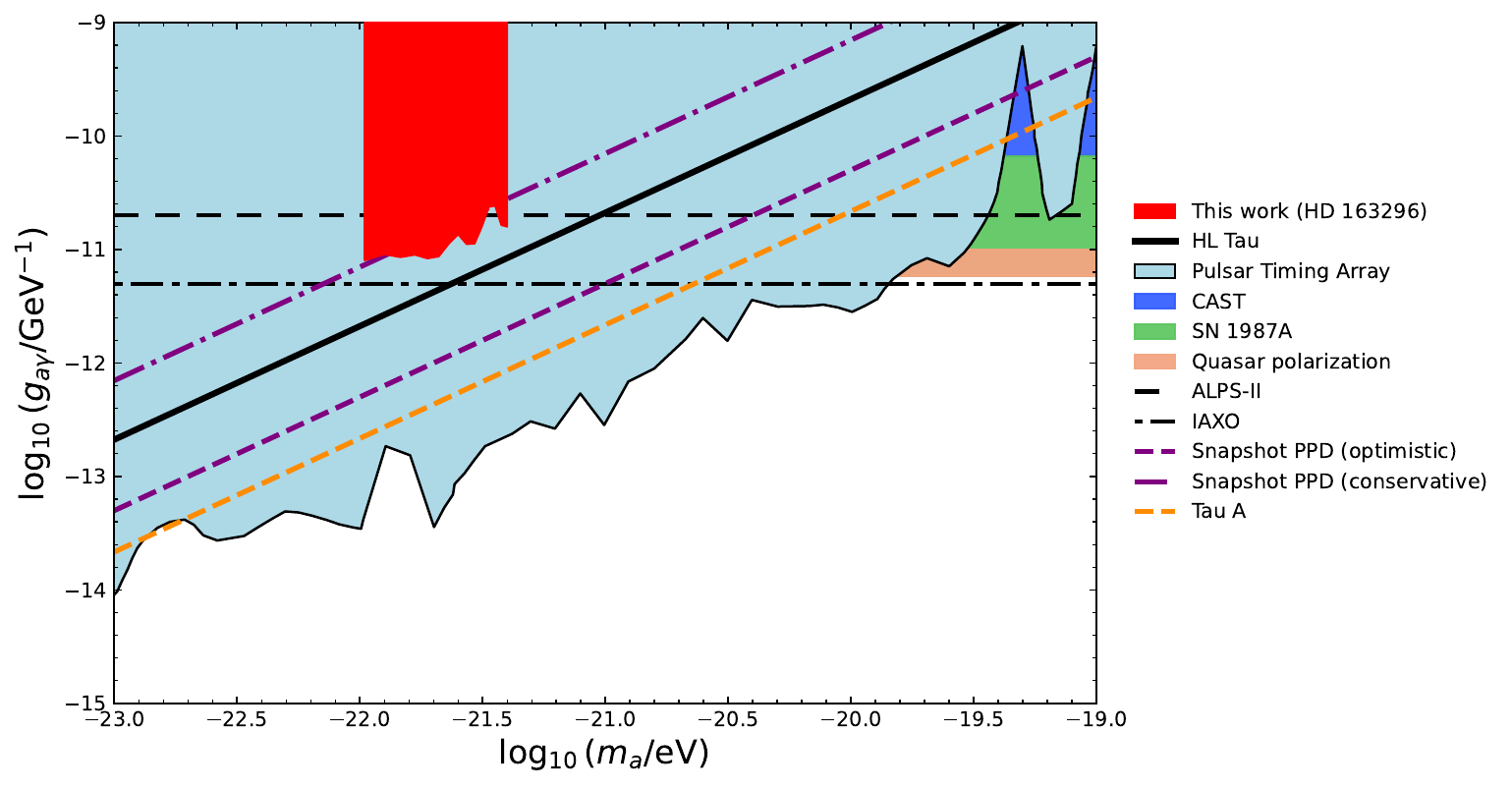}
\caption{
$95\%$ CL upper limit on the axion–photon coupling constant derived from the HD~163296 dataset. 
The accessible frequency range corresponds to oscillation periods between $\sim$100 and 400~days, or ALP masses $m_a \sim 10^{-22}$–$10^{-21}$~eV. 
The red shaded region represents the constraint obtained in this work from the HD~163296 polarization-angle time series.
The purple dash-dotted line and dashed line denote the optimistic and conservative upper bounds from PPD polarization reported by \cite{2019PhRvL.122s1101F}, respectively.
The black solid line shows the constraint derived from multi-epoch ALMA polarization observations of the protoplanetary disk HL~Tau by Davydov and Libanov \citep{2024PhRvD.110j3022D}.
The orange dashed line indicates the constraint obtained from Tau~A polarization measurements \citep{2024PhRvD.110f3013A}.}
The blue, green, orange, and light blue shaded regions are excluded by laboratory (CAST \citep{2017NatPh..13..584C}) and astrophysical observations (SN 1987A \citep{2015JCAP...02..006P}, quasar polarization \citep{2012JCAP...07..041P} and Pulsar Timing Array \citep{2024arXiv241202229X}).
\label{fig:hd163296_limit}
\end{figure}

Figure~\ref{fig:evpa_HD163296} shows the time series of the disk-averaged polarization angle for HD~163296,
derived from the $Q_{\phi}$ and $U_{\phi}$ images at each epoch.
Across all epochs, the representative polarization angles are mutually consistent
within the measurement uncertainties, and no statistically significant temporal
variability is detected.
The observed scatter in the polarization angle measurements can be fully accounted
for by the estimated uncertainties. As a result, the present data provide no evidence for polarization-angle variations
arising from exotic physics such as axion-induced rotation of polarization plane.
Probing subtler temporal variations or coherent periodic signals will require
observations with higher cadence and a longer temporal baseline than available in
the current dataset.
In addition, combining polarization time-series observations of multiple sources
spanning different physical environments will be essential to distinguish
propagation effects, such as axion-induced polarization rotation, from
source-intrinsic astrophysical variability. Using this polarization angle time series, we demonstrate the feasibility of
constraining ALP-induced rotation of polarization plane with PPDs.
Although the number of epochs is insufficient to perform a detection search,
we can place a statistically meaningful 95\% confidence-level upper limit on the
amplitude of possible polarization-angle modulations.
We applied the sinusoidal fitting and Monte Carlo procedure described in Section III to the HD 163296 polarization angle time series. The temporal sampling of this dataset spans $T_{\rm span}\simeq399$ days with a median interval of $\Delta t_{\rm med}\simeq49$ days, corresponding to an accessible frequency range of

\begin{equation}
f_{\rm min}\simeq \frac{1}{T_{\rm span}},
\qquad
f_{\rm max}\sim \frac{1}{2\Delta t_{\rm med}},
\end{equation}

where the lower limit follows from the requirement that at least one full oscillation period be contained within the observing baseline, while the upper limit is approximately set by the effective Nyquist frequency associated with the sampling cadence. These limits correspond to oscillation periods of approximately 100--400 days, which translate through $m_a = 2\pi f$ into an ALP mass range of approximately $10^{-22}$--$10^{-21}\,\mathrm{eV}$.
The sinusoidal search was performed over a grid of trial
frequencies spanning this interval.
The frequency range was sampled from
$f_{\rm min}=1/T_{\rm span}$ to
$f_{\rm max}=1/(2\Delta t_{\rm med})$
with a step size
$\Delta f=(1/T_{\rm span})/5$,
corresponding to an oversampling factor of five relative
to the independent Fourier spacing.
No significant periodic modulation was detected at any trial frequency. 
The 95\% confidence-level upper limits on the modulation amplitude, $A_{95}(f)$, are translated into constraints on the axion--photon coupling via Equation~ \ref{eq:gagg_limit_rho03}, producing the exclusion curve shown in Figure~\ref{fig:hd163296_limit}.

Adopting the most conservative limit, $A_{95}=6^\circ$, we obtain
\begin{equation}
g_{a\gamma} \lesssim 7.5 \times 10^{-12}\,(m_a / 10^{-22}\,\mathrm{eV})\,\mathrm{GeV}^{-1}.
\end{equation}
The modest weakening of the exclusion curve around
$m_a \sim 5\times10^{-22}\,\mathrm{eV}$
does not reflect any physical feature of the ALP model.
Instead, it arises from the finite temporal sampling and
cadence of the observations, which reduce the sensitivity
to oscillation periods comparable to the characteristic
gaps in the time series.

We note that polarization-angle constraints derived from PPD observations
can depend on how the representative angle and its uncertainty are defined.
For an approximately axisymmetric disk, the central fitted value of the
polarization-angle distribution is primarily determined by the geometric symmetry
of the system and is therefore expected to be largely insensitive to
multiple-scattering effects.
In contrast, the dispersion of the polarization-angle distribution can be
broadened by radiative-transfer complexities and multiple scattering
within the disk atmosphere.
In the previous analysis of PPD polarization \citep{2019PhRvL.122s1101F},
the bound derived from the fitted central value of the polarization angle
was interpreted as an optimistic limit,
while the constraint obtained from the distribution width
was regarded as a conservative bound.
In the present work, the statistical uncertainty of the polarization angle
is evaluated directly from the observed dispersion of the angle distribution.
The resulting limit is therefore directly comparable to the conservative bound.
We find that our constraint is broadly consistent with the conservative limit
reported in the previous study, indicating that our result is not artificially tightened
by idealized assumptions and remains robust against plausible disk-scattering effects.

\subsection{Forecasts and limitations for future polarization angle precision}
\label{sec:forcast}

Based on the formalism introduced in Section~\ref{sec:adm}, 
the uncertainty of the polarization angle follows Equation~(\ref{eq:evpa_block_sigma_chi}), 
which is valid for arbitrary $\sigma_{Q_\phi}$ and $\sigma_{U_\phi}$. 
In the case of $\sigma_{Q_\phi} = \sigma_{U_\phi} \equiv \sigma$, 
it reduces to
\begin{equation}
\sigma_\chi \;\approx\; \frac{1}{2}\frac{\sigma}{P}.
\end{equation}
where $P$ is the polarized intensity defined in
Equation~(\ref{eq:polarized_intensity}), and $\sigma$ denotes the rms noise of the Stokes $Q_\phi$ and $U_\phi$ maps.
For a given target, if $N$ independent pixels are available, the statistical uncertainty improves as
\begin{equation}
\sigma_\chi(N) \;\approx\; \frac{1}{2}\frac{1}{(P/\sigma)\sqrt{N}}
= \frac{1}{2 \mathrm{SNR}  \sqrt{N}},
\label{eq:evpa_precision}
\end{equation}
where $\mathrm{SNR} \equiv P/\sigma$ represents the signal-to-noise ratio per pixel. Here, $N$ denotes the number of statistically independent pixels contributing to the measurement. 
In the forecast, we assume an idealized situation in which all contributing pixels have comparable 
signal-to-noise ratios and probe the same underlying polarization signal. 
In real observations, the signal-to-noise ratio generally varies across the disk, so the achievable polarization-angle precision may be somewhat worse than the simple scaling given by Equation (\ref{eq:evpa_precision}).

Equation~\eqref{eq:evpa_precision} allows us to forecast the observational requirements to achieve a given polarization angle precision.
Specifically, the condition can be written as
\begin{equation}
\mathrm{SNR} \sqrt{N} \;\gtrsim\; \frac{1}{2 \,\sigma_\chi}.
\label{eq:snr}
\end{equation}
For a target precision of $\sigma_\chi=0.1^\circ$, this requires $\mathrm{SNR} \sqrt{N}\gtrsim 286$.
Equation~(\ref{eq:snr}) describes only the statistical component of the polarization-angle uncertainty. Instrumental polarization, polarization crosstalk, uncertainties in the Mueller-matrix calibration, temporal variations in the telescope and instrument response, observing-condition variations, and data-reduction effects may introduce systematic errors that do not necessarily decrease as $N^{-1/2}$.

Van Holstein et al.~\cite{2020A&A...633A..64V} estimated polarization-angle uncertainties of approximately $0.13^\circ$--$0.21^\circ$ for an assumed strongly polarized ($30\%$) circumstellar disk and approximately $0.86^\circ$--$3.0^\circ$ for an assumed weakly polarized ($1\%$) source. These values are idealized, model-based estimates obtained by propagating the uncertainties in the SPHERE/IRDIS Mueller-matrix calibration for the assumed source polarization fractions. They illustrate the strong dependence of the model-estimated angle accuracy on the polarization fraction and instrumental configuration, but do not represent total polarization-angle uncertainties demonstrated in science observations or a demonstrated long-term epoch-to-epoch systematic floor. We therefore use them only as a literature-based order-of-magnitude scale for the instrumental calibration uncertainty.

We therefore regard $\sigma_\chi \sim 0.1^\circ$ as an optimistic but literature-motivated order-of-magnitude benchmark for observations of strongly polarized disks, rather than as a demonstrated total polarization-angle accuracy in science observations or a demonstrated long-term epoch-to-epoch systematic floor for SPHERE/IRDIS. In particular, it has not yet been demonstrated that the total epoch-to-epoch uncertainty, including instrumental calibration, temporal stability, observing-condition variations, data-reduction effects, and source-intrinsic variability, can be controlled at this level over a long-term monitoring campaign. The forecast below is therefore conditional on achieving a total polarization-angle uncertainty of approximately $0.1^\circ$, including these systematic effects.

\begin{figure}[t]
\centering
\includegraphics[width=0.65\textwidth]{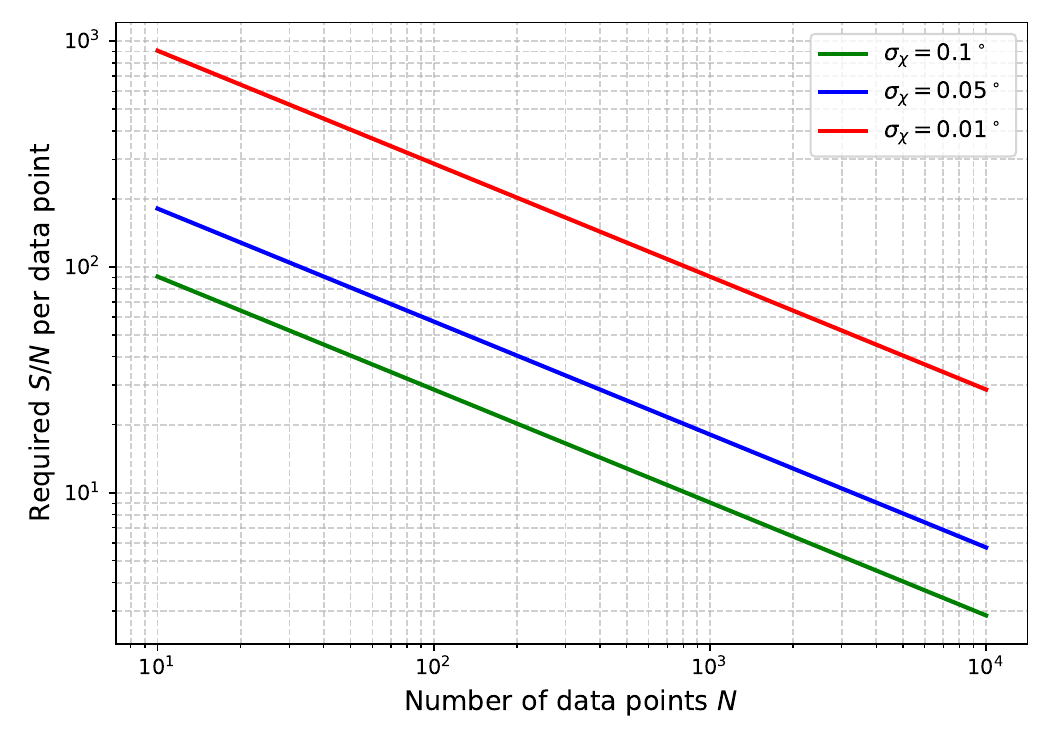}

\caption{Required signal-to-noise ratio per measurement (S/N) as a function of the number of independent measurements $N$ for target statistical polarization-angle precisions of $\sigma_\chi=0.1^\circ$, $0.05^\circ$, and $0.01^\circ$. The figure illustrates the statistical trade-off between observational depth and the number of independent measurements.} 

\label{fig:evpa_forecast}
\end{figure}

These scaling relations are illustrated in Fig.~\ref{fig:evpa_forecast}, where the required signal-to-noise ratio per pixel is shown as a function of the number of independent pixels $N$ for different target precisions. The figure illustrates only the statistical trade-off between the observational depth (SNR per pixel) and the spatial sampling ($N$). It does not demonstrate that the corresponding total polarization-angle uncertainties are achievable, because systematic effects may impose an epoch-to-epoch uncertainty floor.

Here we aimed to obtain a statistically reliable estimate of the upper limit on the polarization angle modulation amplitude $A_{\rm lim}$ under a setup that closely reflects actual observations.
Specifically, we assumed an idealized monitoring scenario in which a representative target in Taurus is observed for $\sim150$ nights per year over a 10-year period.
Using the statistically estimated values of $\sigma_\chi(N)$ for such a single target, we generated noise-only time-series realizations matched to this cadence.
Through this procedure, we evaluated the expected distribution of fitted amplitudes and hence the corresponding sensitivity to the ALP–photon coupling constant.
We emphasize that this monitoring scenario is intended as an illustrative optimistic case rather than a specific observing program. In practice, seasonal visibility, weather conditions, and telescope scheduling constraints may reduce the achievable cadence, thereby weakening the projected sensitivity.

\begin{figure}[t]
\centering
\includegraphics[width=0.8\textwidth]{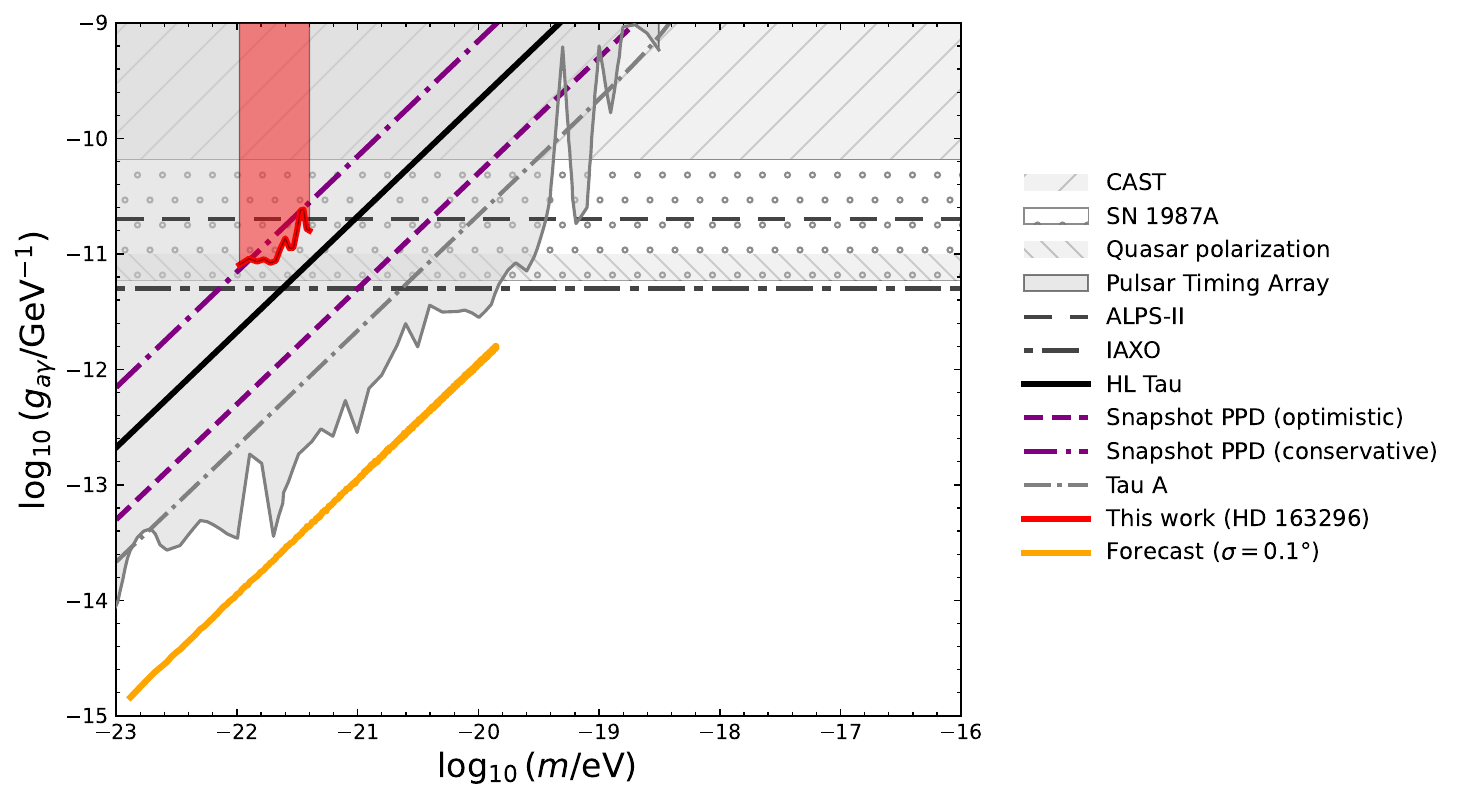}
\caption{Conditional forecast of the 95\% confidence-level (CL) sensitivity to the axion--photon coupling $g_{a\gamma}$ from long-term monitoring of protoplanetary disks, assuming a total polarization-angle uncertainty of $\sigma_\chi=0.1^\circ$, including systematic effects. The orange solid line shows the corresponding expected sensitivity. Other shaded regions and reference limits are the same as in Fig.~\ref{fig:hd163296_limit}. 
}
\label{fig:upperlimits_forecast}
\end{figure}
The accessible ALP mass range is determined by the temporal sampling of the assumed monitoring campaign. The lowest accessible masses correspond to oscillation periods comparable to the total monitoring baseline, since periods much longer than the observing span cannot be reliably distinguished from slow secular trends. Conversely, the highest accessible masses are limited by the Nyquist frequency associated with the observing cadence. For the assumed cadence of approximately 150 observations per year over a 10-year baseline, the corresponding mass range is approximately $10^{-23} \lesssim m_a \lesssim 10^{-20}$ eV.

Under the conditional assumption that a total polarization-angle uncertainty of $\sigma_\chi=0.1^\circ$, including systematic effects, can be achieved, the projected sensitivity shown in Fig.~\ref{fig:upperlimits_forecast} could rival or surpass several existing astrophysical constraints. As discussed above, this value is adopted as an optimistic, literature-motivated order-of-magnitude benchmark for strongly polarized disk observations; it is not a demonstrated long-term epoch-to-epoch systematic floor. The forecast therefore illustrates the potential scientific reach of high-stability polarimetric observations under this uncertainty assumption, rather than the expected performance of the current SPHERE/IRDIS system.

\section{Discussion}
\label{sec:discussion}
\subsection{Outlook}
In this study, we focus on optically thick PPDs, in which multiple scattering by dust grains plays a crucial role. As a result of this multiple scattering, the spatial distribution of polarization angles is broadened to scales of several tens of degrees, and an intrinsic uncertainty at the level of a few degrees is introduced in the determination of the representative polarization angle.
In contrast, for optically thin debris disks \citep[e.g.,][]{2024ApJ...977..277S}, where no additional physical mechanisms break axisymmetry, multiple scattering is expected to be negligible and single scattering dominates. Under such idealized conditions, the polarization angle distribution approaches a delta-function–like profile if we neglect instrumental errors, enabling an extremely precise determination of the intrinsic polarization angle. Debris disks therefore provide particularly clean targets for probing minute polarization rotations. However, debris disks are generally faint in scattered light, making high-precision polarimetric measurements challenging with current ground-based telescopes.
In addition to target selection, the achievable polarization angle precision itself is a key factor. The attainable precision depends not only on the signal-to-noise ratio, but also strongly on the number of independent measurements $N$, as expressed by Equation~(\ref{eq:snr}). Furthermore, broader temporal sampling across multiple disks allows the search to probe a wider range of ALP masses, highlighting the importance of observing diverse targets over extended periods. Since a single target typically provides only a limited number of epochs, combining time-series data from a large number of disks offers a practical way to enhance sensitivity and extend temporal coverage. Consequently, a strategy that integrates long-term observations of many disks is expected to maximize the overall sensitivity to ALP-induced rotation of polarization plane. The methodology and limitations of such joint analyses across multiple disks are discussed in detail in Section~\ref{subsec:joint_limit}.

As discussed in Section~\ref{sec:forcast}, the $\sigma_\chi\sim0.1^\circ$ precision adopted in our forecast is an optimistic, literature-motivated order-of-magnitude benchmark, rather than a demonstrated long-term performance of current instruments. While Mueller-matrix modeling indicates polarization-angle accuracies approaching this order for strongly polarized disks, the total epoch-to-epoch systematic uncertainty relevant to a multi-year monitoring campaign has not yet been established at this level. Reaching the conditional forecast would require the combined effects of instrumental calibration, temporal stability, observing-condition variations, data reduction, and source-intrinsic variability to be controlled such that the total uncertainty remains approximately $0.1^\circ$. Favorable targets, such as optically thin debris disks with minimal astrophysical polarization-angle dispersion, may help approach this regime.

\subsection{Limitations of Joint Time-Series Analysis Across Multiple PPDs}
\label{subsec:joint_limit}

A natural strategy to boost sensitivity is to combine multi-epoch polarization time series from different PPDs. 
Several approaches to such a joint analysis are possible,
including methods that search for a common oscillation period without
assuming a shared phase \citep{2019JCAP...02..059I}. In this section, however, we focus
on coherent time-domain analyses, which can provide the highest
sensitivity when the ALP field remains phase-coherent across the
targets.
However, such a joint coherent analysis is subject to two fundamental limitations: (i) the assumption of a common oscillation phase and field coherence, and (ii) distance-induced phase uncertainty.
These two limitations are qualitatively different in nature:
the former reflects an intrinsic phase freedom of the ALP field that
cannot be constrained by independent observations, whereas the latter
arises from geometric distance uncertainties and can, in principle,
be mitigated by improved astrometric measurements.

\paragraph*{(i) Common oscillation phase and field coherence.}
For coherent time-domain stacking, a coherent stack requires that all targets share not only the same ALP mass (and hence the same oscillation frequency $\omega = m_a$), but also a common oscillation \emph{phase} of the ALP field. 
In principle, each source could have an independent emission phase $\delta_{\rm emit}$, which would introduce many additional free parameters in a combined fit. Since this intrinsic emission phase cannot be determined by other observations, it must be treated as a free parameter in a time-domain analysis.
Importantly, within a single coherence domain of the ALP field, whose characteristic size is set by the de~Broglie wavelength, the field phase can be regarded as spatially uniform:
\begin{equation}
\lambda_{\rm coh}
\simeq 40~\mathrm{pc}
\left(\frac{10^{-22}~\mathrm{eV}}{m_a}\right),
\end{equation}
which corresponds to an oscillation period of approximately
$\sim $ year.
This greatly reduces the parameter space, allowing all targets within the coherence length to be modeled with a single common phase. 
Objects separated by more than the coherence length should instead be analyzed independently or combined only incoherently (e.g., at the power spectrum level).

\paragraph*{(ii) Distance-induced phase uncertainty.}
In contrast to the intrinsic emission phase discussed above,
even within a shared coherence domain, accurate phase alignment requires
correcting for the geometric light-travel delay between each source and
the observer. This effect therefore does not represent an intrinsic free parameter of the model, but rather a source of phase uncertainty that can be reduced with sufficiently precise distance measurements.

Denoting the source distance by $d$ with uncertainty $\sigma_d$, the relevant phase contribution is
\begin{equation}
\phi_d = \frac{m_a}{2}(t_{\rm obs}-t_{\rm emit}) = \frac{m_a d}{2c},
\end{equation}
with an associated uncertainty
\begin{equation}
\delta\phi_d = \frac{m_a}{2c}\,\delta d, 
\qquad 
\sigma_{\phi_d} = \frac{m_a}{2c}\,\sigma_d.
\end{equation}

For Gaia-like astrometric precision of $\sigma_d \sim 0.1$~pc, one finds $\sigma_{\phi_d} \simeq 0.75$ rad at $m_a = 10^{-22}$~eV, but $\sigma_{\phi_d} \simeq 7.5$ rad at $m_a = 10^{-21}$~eV, implying severe coherence loss at higher masses. 
A practical criterion for maintaining coherence is $\sigma_{\phi_d}\lesssim 1$ rad, i.e.,
\begin{equation}
\sigma_d \lesssim \frac{2c}{m_a}
\simeq 0.13~\mathrm{pc}\,\left(\frac{10^{-22}~\mathrm{eV}}{m_a}\right)
= 0.13~\mathrm{pc}\,\,m_{22}^{-1}.
\end{equation}

\section{Conclusion}
\label{sec:conclusion}

We have presented a dedicated time-series search for axion-induced polarization-angle oscillations using multi-epoch polarization observations of a protoplanetary disk.
Using archival near-infrared polarimetric observations of HD~163296, we performed a consistent multi-epoch analysis of the polarization angle, extending previous single-epoch studies to the time domain.
As a result, the six-epoch polarization-angle time series shows no statistically significant variability beyond the measurement uncertainties.

From the absence of detectable variability, we derived the first constraints on the ALP--photon coupling based on PPD polarization, placing a 95\% confidence-level upper limit of
\begin{equation}
g_{a\gamma} \lesssim 7.5 \times 10^{-12}
\left( \frac{m_a}{10^{-22}\,\mathrm{eV}} \right)
\mathrm{GeV}^{-1}.
\end{equation}
Although this constraint is not yet competitive with the strongest existing astrophysical limits, it establishes PPD polarization variability as a viable and previously unexplored probe of ultralight ALPs.

A key limitation of the current dataset is the relatively large polarization angle uncertainty, which is partly driven by multiple scattering in the optically thick disk.
This effect broadens the intrinsic polarization angle distribution and introduces additional dispersion in the representative polarization angle, limiting sensitivity to small rotation of polarization plane.

Nevertheless, our Monte Carlo forecasts demonstrate the potential improvement under an explicitly conditional uncertainty assumption. In particular, if future observations can achieve a total polarization-angle uncertainty of $\sigma_\chi\sim0.1^\circ$, including systematic effects, observations of favorable targets such as optically thin debris disks could provide sensitivity to ALP-induced polarization-plane rotation that rivals or surpasses existing astrophysical constraints. Here, $\sigma_\chi\sim0.1^\circ$ should be understood as an optimistic, literature-motivated order-of-magnitude benchmark rather than a demonstrated long-term systematic floor.

With a dense long-term monitoring strategy, consisting of $\sim$150 observing epochs per year over a $\sim$10-year baseline, PPD and debris-disk polarization could probe ALP masses in the $10^{-23}$--$10^{-20}$~eV range.
This mass window is complementary to those targeted by pulsar timing arrays, CMB polarization, and other astrophysical and cosmological searches.

As such, time-domain polarimetry of circumstellar disks offers an independent and conceptually clean avenue for testing ultralight dark matter, grounded in well-understood polarization geometries.
Future high-sensitivity, multi-epoch polarimetric observations will therefore open a promising new frontier in the search for axion-like dark matter.

\begin{acknowledgments}
We thank Bin Ren for making the reduced VLT/SPHERE images of HD 163296 publicly available.
We are grateful to Yuji Chinone, Yasuo Doi, Jun Hashimoto, Akimasa Kataoka and Naoya Kitade for fruitful discussions.
This work was supported by JST SPRING, Grant Number JPMJSP2108
and by JSPS KAKENHI Grant Number JP23K03424.
RT was also supported by  JSPS KAKENHI (Grant Numbers JP25K07351, JP25K01049).
KN was supported by the International Graduate Program for Excellence in Earth-Space Science (IGPEES), the University of Tokyo, and by JSPS KAKENHI Grant Number JP26KJ0864.
\end{acknowledgments}




\bibliographystyle{apsrev4-2}
\bibliography{apssamp}

\end{document}